# Semi-empirical analysis of leptons in gases in crossed electric and magnetic fields, Part II: Transverse compression of muon beams

Malte Hildebrandt[1, a)], Robert E. Robson[2] and Nathan Garland[3]

[1)] *Paul Scherrer Institut, 5232 Villigen PSI, Switzerland*
[2)] *James Cook University, Townsville, QLD 4811, Australia*
[3)] *Griffith University, Nathan, QLD, Australia*

[a)] *malte.hildebrandt@psi.ch*

(Dated: 3 August 2023)

This article employs fluid equations to analyse muon beams in gases subject to crossed electric and magnetic fields, focussing in particular on a scheme proposed by D. Taqqu in 2006, whereby transverse compression of the beam is achieved by creating a density gradient in the gas. A general criterion for maximising beam compression, derived from first principles, is then applied to determine optimal experimental conditions for $\mu^+$ in helium gas. Although the calculations require input of transport data for ($\mu^+$, *He*), which are generally unavailable, this issue is circumvented by "aliasing" ($\mu^+$, *He*) with ($H^+$, *He*), for which transport coefficient data are available.

## 1. Introduction

*1.1 MTT and compression of muon beams*

In the previous article[1] (referred to as I below) we showed how the fluid equations of momentum transfer theory[2,3] (MTT) could be used to estimate the Lorentz angle for electrons in helium gas in crossed electric and magnetic fields, using experimental swarm data in an electric field only, symbolically:

(*e⁻, He*) swarm data for *E* field only + MTT ➔ Lorentz angle for crossed *E × B* fields

Thus, the equations of MTT acted as an intermediary for processing empirical data obtained in one particular situation in order to obtain information about another, more general case. In addition to generating results in good agreement with simulation techniques[4,5,6], the method has the advantage of being both physically transparent and self-contained.

The present article deals with a beam of positive muons $\mu^+$ in helium gas, subject to crossed electric and magnetic fields. Again, calculations are carried out using MTT, though the emphasis and aims are quite different from I. Thus, we consider transverse compression of the muon beam through imposition of a density gradient in the gas, as proposed by Taqqu[7], and derive criteria for generating the gas density profile which maximises the compression. In contrast to I, the Lorentz angle is specified at the outset, while MTT acts as an intermediary for generating the required optimal experimental conditions, i.e., symbolically

<p align="center">Lorentz angle + MTT → Optimal density gradient in the gas</p>

An additional component of empiricism is required, since the procedure requires input of $(\mu^+, He)$ transport data which is, however, unavailable. A similar problem also arises in simulations, which require input of unknown $(\mu^+, He)$ scattering cross sections. However, just as Taqqu[7] argued that known $(H^+, He)$ that cross sections could adapted to the muon problem for simulation purposes, we adapt or "alias" known $(H^+, He)$ swarm experimental transport data[8,9] to $(\mu^+, He)$. Such a procedure has already been discussed in the context of MTT for muons in an $E$ field only[10], and the present paper extends the method to crossed $E$ and $B$ fields.

This article, like I, provides what we believe to be an efficient, self-contained and physically transparent procedure, over which the user has total control, thereby offering an alternative, complementary approach to simulation packages (e.g., GEANT[4], Magboltz[5,6]). In addition, the prescription we provide for optimising transverse compression can be incorporated into both simulations and fluid analysis, and is directly applicable to experiment.

*1.2 Outline of this article*

This article involves the conjunction of two distinct fields, which come under the general headings of beam physics and transport analysis respectively. Readers familiar with one area may not be familiar with the other, and we therefore commence with a brief review of muon beams in Section 2, followed by a summary in Section 3 of the fluid equations of momentum transfer theory (MTT). Section 4 outlines how this theory is to be applied to beam analysis, starting with derivation of an expression for stopping power[11]. In Section 5 Taqqu's proposal[7] for beam compression in a gas with a spatially varying density is then explored, and we obtain criteria for maximum transverse compression. This "maximum compression theorem" is then applied in Section 6 to obtain optimal conditions for a $(\mu^+, He)$ compression cell. Section 7 concludes the article with a discussion of the results.

## 2. Reasons for studying muon beams

The fundamental particle muon plays an important role in low energy, precision particle physics and material science[12,13]. In particular experiments with positive muons ($\mu^+$) and muonium atoms ($\mu^+e^-$) offer promising possibilities for tests of fundamental symmetries and searches for physics beyond the standard model of particle physics, e.g., the measurement of the muon anomalous magnetic moment (g-2)[14,15], the search for a muon electric dipole moment[15], muonium spectroscopy[16,17] or muonium gravity measurements[18,19]. The next generation of these precision particle physics experiments or muon Spin Resonance research methods in material science require high-intensity low energy muon beams with very small phase space, i.e. with small transverse size and energy spread.

Nowadays, standard muon beams available at meson facilities[19,20] are produced by pion decay and due to the large area of the production target and the large capture acceptance of the beam line elements of the secondary beamline the beams have relatively poor phase space quality and high energy. Common phase space compression schemes based on stochastic cooling, electron cooling or muon energy moderation in materials are not applicable due to the short muon life time (~2.2 μs) or low cooling efficiencies ($<10^{-4}$).

Several years ago, a new compression scheme[7] was proposed and a novel device[21] that reduces the full phase space of a $\mu^+$ beam by 10 orders of magnitudes with an efficiency of $10^{-3}$ is currently under development at the Paul Scherrer Institute. The applied method[22,23] moderates a positive muon beam in a few mbar cryogenic helium gas while simultaneously compressing its beam spot in an $E \times B$ fields.

## 3. Fluid equations in beams and swarm experiments

*3.1 A brief resume of fluid modelling*

Consider a beam or "swarm" of low-energy charged particles of mass $m$, number density $n$, interacting with a target gas (e.g. helium) of far more numerous atoms or molecules of mass $m_0$, number density $n_0 >> n$, as shown schematically in Fig. 1:

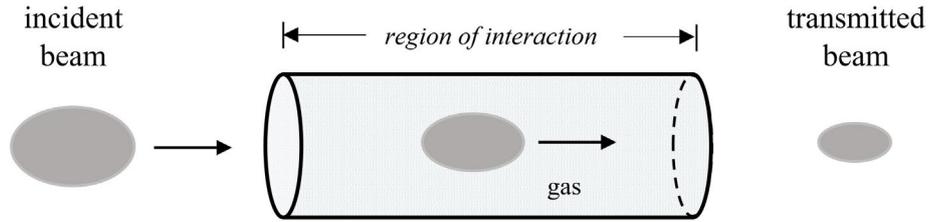

**Fig. 1**. A low density beam of particles interacts with a target gas of known properties. Electric and magnetic fields (not shown) may be applied in the region of interaction to achieve the desired acceleration and dispersion (longitudinal and/or lateral) of the beam.

In a time-of-flight experiment[24], the time taken for the centre-of-mass of the swarm to move a known distance under the influence of external electric and magnetic fields, plus the dispersion of particles about the centre-of-mass, are measured to yield the drift velocity and longitudinal and transverse diffusion coefficients respectively. These quantities can be calculated theoretically by either:

(i) Kinetic theory[25,26], in which one first solves Boltzmann's equation for the swarm velocity distribution function $f(\mathbf{v})$ followed by appropriate averaging over $\mathbf{v}$ to obtain the average particle velocity $\overline{\mathbf{v}}$, average kinetic energy $\overline{1/2 m v^2}$, and so on; or

(ii) Fluid theory[2,3], where the averages are obtained directly as solutions of macroscopic conservation equations for momentum, energy and so on, without the need to find $f$ first. Of particular interest are the expressions for the average rate of transfer of these properties in collisions.

Although (ii) provides a much faster way of finding the physically measureable quantities, this "short cut" comes at the price of reduced accuracy, since some approximation is inevitable. Thus, whereas kinetic theory typically yields physically measurable quantities accurate to 1% or better, the fluid equations used here, based on momentum-transfer theory[2], furnish the required averages accurate to around 10%.

Note that the neutral gas, whose properties are specified by a subscript 0, is at rest and in equilibrium at temperature $T_0$, so that $\overline{\mathbf{v}}_0 = 0$, $\overline{\frac{1}{2} m_0 v_0^2} = \frac{3}{2} k T_0$.

Fundamental to the calculations is the collision frequency for momentum transfer:

$$v_m(\varepsilon) = n_0 \sqrt{\frac{2\varepsilon}{\mu_r}} \sigma_m(\varepsilon) \qquad (1)$$

where

$$\sigma_m(\varepsilon) = 2\pi \int_0^\pi \sigma(\varepsilon, \chi)[1 - \cos\chi] \sin\chi \, d\chi \qquad (2)$$

is the momentum transfer cross section, $\varepsilon$ is the centre-of-mass energy, $\chi$ is the scattering angle in the centre of mass frame, $\sigma(\varepsilon, \chi)$ is the differential cross section, and

$$\mu_r = mm_0 / (m + m_0) \qquad (3)$$

is the reduced mass. The following mass ratios also appear in what follows:

$$M = m / (m + m_0), \quad M_0 = m_0 / (m + m_0)$$

*3.2 The collision terms*

The energy-dependence of the cross section is determined by the nature of the charged particle-atom interaction potential, e.g., for an inverse-fourth power law potential, $\sigma_m(\varepsilon) \sim \varepsilon^{-1/2}$, and by Eqn (1), $v_m$ is a constant, independent of energy (Maxwell model[2,3,25,26]). In this special case, the collision terms in the fluid equations can be represented exactly, but otherwise an approximation is required. MTT consists in assuming that fluid equations of the same *mathematical form* as the Maxwell model can be employed in more general circumstances. In particular, the mean collisional transfer rates of momentum and energy in elastic collisions are given by:

$$\overline{\left(\frac{\partial m\mathbf{v}}{\partial t}\right)_{col}} \approx -\mu_r v_m(\overline{\varepsilon}) \, \overline{\mathbf{v}} \qquad (4)$$

and

$$\overline{\left(\frac{\partial m v^2 / 2}{\partial t}\right)_{col}} \approx -2M v_m(\overline{\varepsilon}) \, (\overline{\varepsilon} - \frac{3}{2} kT_0) \qquad (5)$$

respectively. For the Maxwell model, these expressions are exact, and otherwise, the approximations are better for collision frequencies which vary slowly with energy.

For gases whose constituent atoms and molecules have internal states *I* which may be altered through *inelastic collisions* and Eqn (15) must be modified by the inclusion of an additional term[2,3]. Thus, in general,

$$\overline{\frac{\partial m v^2 / 2}{\partial t}}\bigg)_{col} \approx -2M\nu_m(\overline{\varepsilon})(\overline{\varepsilon} - \frac{3}{2}kT_0) - M_0 \sum_I \varepsilon_I \left(\overrightarrow{\nu_I} - \overline{\overrightarrow{\nu_I}}\right) \quad (6)$$

where

$$\vec{\nu}_I \equiv \overrightarrow{\nu_I(\varepsilon)} = n_0 \overline{\sqrt{\frac{2\varepsilon}{\mu_r}} \vec{\sigma}_I(\varepsilon)} \quad (7)$$

denotes the average excitation collision frequency characterised by cross section $\vec{\sigma}_I(\varepsilon)$ with threshold energy $\varepsilon_I$, and $\overleftarrow{\nu}_I$ denotes the corresponding quantity for the inverse process, i.e., de-excitation. Well above threshold, $\overline{\varepsilon} \gg \varepsilon_I$, inelastic cross sections (and therefore inelastic collision frequencies) may vary sufficiently smoothly, making it reasonable to assume that

$$\overrightarrow{\nu_I(\varepsilon)} \approx \vec{\nu}_I(\overline{\varepsilon}) = n_0 \sqrt{\frac{2\overline{\varepsilon}}{\mu_r}} \vec{\sigma}_I(\overline{\varepsilon}) \quad (8)$$

At high energies, $\overline{\varepsilon} \gg \frac{3}{2}kT_0$, the thermal motion of the gas molecules can be neglected (cold gas), and inverse collisions may be neglected, $\overleftarrow{\nu}_I \to 0$. Thus the high energy limit of (6) is

$$\overline{\frac{\partial m v^2 / 2}{\partial t}}\bigg)_{col} \approx -2M\nu_m(\overline{\varepsilon})\overline{\varepsilon} - M_0 \sum_I \varepsilon_I \vec{\nu}_I(\overline{\varepsilon}) \quad (9)$$

In addition for a cold gas $\overline{\varepsilon} \approx \frac{1}{2}\mu_r \overline{v^2}$, and therefore $\sqrt{\frac{2\overline{\varepsilon}}{\mu_r}} \approx \sqrt{\overline{v^2}} \equiv v_{rms}$. Then (9) can be written in the alternative form:

$$\overline{\frac{\partial m v^2 / 2}{\partial t}}\bigg)_{col} \approx -v_{rms}\left[2M\sigma_m(\overline{\varepsilon})\overline{\varepsilon} + M_0 \sum_I \varepsilon_I \vec{\sigma}_I(\overline{\varepsilon})\right] \quad (10)$$

This expression will be used in the derivation of an expression for stopping power of muons in a gas in Section 4.1 below.

3.3 *Balance equations for energy and momentum*

As a first approximation we consider a spatially uniform swarm and also neglect any time variation. Using (4) and (6), we find that the fluid equations corresponding to momentum and energy balance in uniform external electric and magnetic fields are[2,3]

$$q(\boldsymbol{E} + \overline{\boldsymbol{v}} \times \boldsymbol{B}) \approx \mu_r \nu_m(\overline{\varepsilon}) \overline{\boldsymbol{v}} \quad (11)$$

and

$$q\boldsymbol{E} \cdot \overline{\boldsymbol{v}} \approx \nu_e(\overline{\varepsilon})(\overline{\varepsilon} - \frac{3}{2}kT_0 + \Omega) \quad (12)$$

respectively, where

$$\Omega \equiv M_0 \sum_I \varepsilon_I \left(\overline{\vec{v}_I} - \overline{\overline{v}_I}\right) / v_e(\overline{\varepsilon}) \tag{13}$$

and

$$v_e(\overline{\varepsilon}) \equiv 2M v_m(\overline{\varepsilon}) \tag{14}$$

is the average collision frequency for elastic energy transfer. Eqns (11) and (12) may be combined to give

$$\overline{\varepsilon} = \frac{3}{2} kT_0 + \frac{1}{2} m_0 \overline{\mathbf{v}}^2 - \Omega(\overline{\varepsilon}) \tag{15}$$

which reduces to the well-known Wannier energy equation[25,26] when inelastic processes are negligible.

In Part I, we did not require the explicit expression (13) for $\Omega$, and neither did we need to know the solution of Eqn (15), only that it shows that the mean C.M. energy can be determined if the average velocity is known, or conversely, i.e., $\overline{\varepsilon} = \overline{\varepsilon}(\overline{\mathbf{v}})$ or $\overline{\mathbf{v}} = \overline{\mathbf{v}}(\overline{\varepsilon})$. Indeed, the same simplifications apply in most of the present paper, and Eqn (10) is required only for the purpose of deriving expressions for stopping power (see Section 4.1 below).

Just as Eqns (11) and (15) provide the framework for fluid analysis of traditional ion and electron swarm experiments, they also provide the basis for modelling compression of muon beams (see Sections 4.2 and 4.3 below). In addition, they also furnish the means of obtaining otherwise unknown properties of muons by "aliasing" proton data for which transport quantities are known[8].

## 4. Stopping power

Before proceeding to the question of muon compression, we make contact with the literature on beams by showing how the MTT formalism yields expressions for stopping power familiar in the beam literature.

By Eqn (10) it follows that the rate of loss of energy per unit distance $x$ along the particle track is

$$\overline{\left.\frac{\partial m\mathbf{v}^2/2}{\partial x}\right)_{col}} = \frac{1}{\mathbf{v}_{rms}} \overline{\left.\frac{\partial m\mathbf{v}^2/2}{\partial t}\right)_{col}} \approx -\left[2M\sigma_m(\overline{\varepsilon})\,\overline{\varepsilon} + M_0 \sum_I \varepsilon_I \vec{\sigma}_I(\overline{\varepsilon})\right] \tag{16}$$

The *stopping power* is defined by

$$S = \frac{1}{n_0} \overline{\frac{\partial m v^2 / 2}{\partial x}}\bigg)_{col} = S^{(el)} + S^{(inel)} \qquad (17)$$

where

$$S^{(el)} = 2M \, \overline{\varepsilon} \, \sigma_m(\overline{\varepsilon}) \qquad (18)$$

and

$$S^{(inel)} = M_0 \sum_I \varepsilon_I \vec{\sigma}_I(\overline{\varepsilon}) \qquad (19)$$

denote the contributions from elastic and inelastic processes respectively. These general expressions can be simplified slightly for lighter particles such as muons through the approximations $M \approx m/m_0$ and $M_0 \approx 1$. In particular Eqn (18) can be written as

$$S^{(el)} = \Delta\overline{\varepsilon} \, \sigma_m(\overline{\varepsilon}) \qquad (20)$$

where

$$\Delta\overline{\varepsilon} = \frac{2m}{m_0} \overline{\varepsilon}$$

is the average kinetic energy lost by a muon in an elastic collision.

Similar expressions for the stopping power can be found in the literature[11].

## 5. Beam compression in external fields

*5.1 Optimal conditions for compression*

The criterion that we shall assume gives optimal beam compression is simply that the mean muon velocity converges in space, i.e.,

$$\nabla \cdot \overline{\mathbf{v}} < 0 \qquad (21)$$

Thus, by calculating an expression for $\overline{\mathbf{v}}$ in terms of the external fields and other parameters, using the balance equations for momentum and energy (11) and (15) respectively, and substituting in (21), we obtain the criterion for convergence in terms of those fields and parameters.

While the externally applied electric and magnetic fields, **E** and **B** respectively, are uniform, the gas number density $n_0$ generally varies with position. If it is assumed that:

- to first order, muon properties are approximately uniform, even though gas properties may vary with position;

- muon density is very low, i.e., $n \ll n_0$, so that muon-muon interactions can be neglected in comparison with muon-neutral molecule collisions; and
- space-charge effects any perturbations of the external fields are negligible

then the momentum and energy balance equations Eqns (11) and (15) for a spatially uniform swarm may be applied, to a first approximation, to the muon beam.

*5.2 Longitudinal beam compression in an electric field*

We start with the simplest case where there is either no magnetic field, or the mean velocity is directed along the magnetic field. In that case the term $\bar{\mathbf{v}} \times \mathbf{B}$ on the left hand side of (11) vanishes, and it follows that

$$\bar{\mathbf{v}} = K\,\mathbf{E} \qquad (22)$$

where

$$K = \frac{q}{\mu_r \nu_m(\bar{\varepsilon})} \qquad (23)$$

is the muon mobility coefficient. Note that the relationship (15) between mean energy $\bar{\varepsilon}$ and mean velocity $\bar{\mathbf{v}}$ remains the same, with or without a magnetic field. Eqns (23) and (15) together enable calculation of mean velocity and mean energy for specified cross sections. Other points to note are that since $\nu_m \sim n_0$ it follows that:

- The mean velocity and mean energy depend upon $E$ and $n_0$ through the *ratio* $E/n_0$ (the "reduced electric field") rather than upon each factor separately; and
- Although (22) is strictly speaking valid only for spatially uniform conditions, we nevertheless apply it to non-uniform situations, resulting from a number density which varies in space, i.e., $n_0 = n_0(\mathbf{r})$. Obviously, this involves some degree of approximation.

Taking the divergence of (22), and using (15) then gives

$$\nabla \cdot \bar{\mathbf{v}} = -\alpha\,\bar{\mathbf{v}} \cdot \frac{1}{n_0}\nabla n_0 \qquad (24)$$

where

$$\alpha \equiv 1 - \frac{m_0 \bar{\mathbf{v}}^2}{1+\Omega'}\frac{\nu_m{}'}{\nu_m} \equiv \frac{2K_d - 1}{K_d} \qquad (25)$$

A prime indicates a derivative with respect to the mean energy, and

$$K_d \equiv \frac{\partial \ln \bar{\mathbf{v}}}{\partial \ln E/n_0} \qquad (26)$$

is the logarithmic differential mobility.

For the special case of a Maxwell model (constant collision frequency), Eqn (25) shows that $\alpha = 1$. Otherwise, in general, given the lack of information about the quantities in right hand side of (25) for muons, we shall simply assume for the purposes of the following discussion that $\alpha$ is of the order of unity.

Firstly, if the gas density gradient $\nabla n_0$ is parallel to $\mathbf{E}$, Eqn (22) then implies that it is also parallel to $\bar{\mathbf{v}}$. Thus it follows from (24) that

$$\nabla \cdot \bar{\mathbf{v}} \sim -E^2 < 0 \tag{27}$$

Thus a muon beam driven through a neutral gas whose density increases in the field direction tends to *converge* or *compress* in the field direction. On the other hand, when the density gradient is perpendicular to the field, and therefore to $\bar{\mathbf{v}}$, Eqn (24) implies that $\nabla \cdot \bar{\mathbf{v}} = 0$. In this case there is a shear in the velocity field rather than compression.

The two cases are portrayed schematically in Fig. 2.

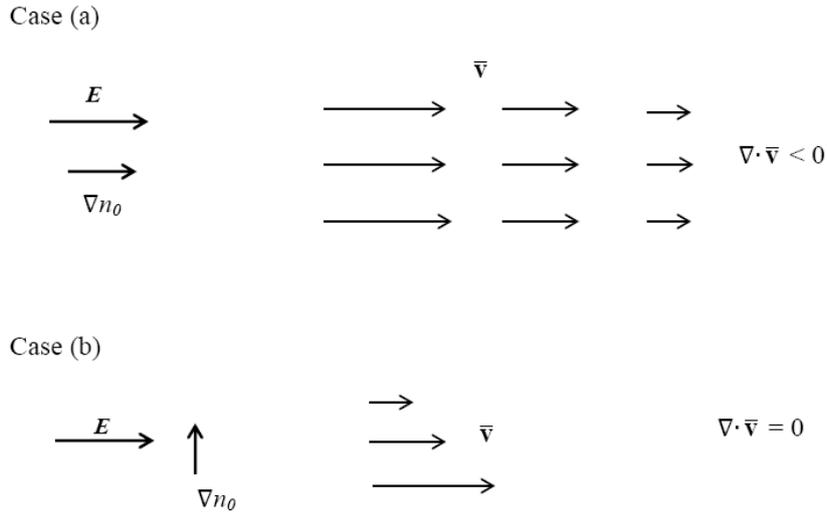

Fig. 2. Longitudinal compression (a) and shear flow (b) of a beam in the presence of a density gradient in the gas.

The most important result here is that maximum longitudinal compression is achieved when the field and density gradient are *parallel*.

*5.3 Transverse beam compression in crossed electric and magnetic fields*

We now consider the case where the beam is compressed in the *transverse* direction by applying a magnetic field in a direction perpendicular to the electric field. For simplicity, and in keeping with the semi-quantitative nature of the discussion, we shall neglect any variation of the collision frequency with energy.

The average drift velocity is found as the solution of (11) as

$$\overline{\mathbf{v}} = \overline{v}\left[\cos\varphi\,\hat{\boldsymbol{E}} + \sin\varphi\,\hat{\boldsymbol{E}}\times\hat{\boldsymbol{B}}\right] \quad (28a)$$

where

$$\overline{v} = KE/\sqrt{1+(KB)^2} \quad (29)$$

is the magnitude of the average velocity, $\varphi$ is the Lorentz angle[24] between $\overline{\mathbf{v}}$ and the direction $\hat{\boldsymbol{E}}$ of the electric field, defined by

$$\tan\varphi = KB = \frac{qB}{\mu_r v_m(\overline{\varepsilon})} \quad (30)$$

and $\hat{\boldsymbol{B}}$ denotes the direction of the magnetic field. Since we are investigating compression in the transverse direction only, we need consider the *direction*

$$\hat{\mathbf{v}} = \overline{\mathbf{v}}/v = \left[\cos\varphi\,\hat{\boldsymbol{E}} + \sin\varphi\,\hat{\boldsymbol{E}}\times\hat{\boldsymbol{B}}\right] \quad (28b)$$

of the mean velocity only, and hence

$$\nabla\cdot\hat{\mathbf{v}} = \frac{\partial\varphi}{\partial n_0}\nabla n_0 \cdot \left[-\sin\varphi\,\hat{\boldsymbol{E}} + \cos\varphi\,\hat{\boldsymbol{E}}\times\hat{\boldsymbol{B}}\right] \quad (31)$$

Note that since $v_m \sim n_0$, Eqn (30) implies that the Lorentz angle decreases with increasing gas number density, i.e.,

$$\partial\varphi/\partial n_0 < 0 \quad (32)$$

Proceeding as in the previous subsection, we note that the right hand side of (31) is a maximum when the density gradient is parallel to the field combination inside the brackets, i.e., when

$$\nabla n_0 = c\left[-\sin\varphi\,\hat{\boldsymbol{E}} + \cos\varphi\,\hat{\boldsymbol{E}}\times\hat{\boldsymbol{B}}\right] \quad (33)$$

where c is some constant. After substituting (33) in (31), and applying (32), we find

$$\nabla\cdot\hat{\mathbf{v}} = c\,\partial\varphi/\partial n_0 < 0 \quad (34)$$

the negative sign on the right hand side indicating convergence. Note further that by (28b) and (33)

$$\hat{\mathbf{v}}\cdot\nabla n_0 = 0 \quad (35)$$

Since $\nabla n_0$ is, by definition, orthogonal to surfaces of constant $n_0$, this implies that the direction $\hat{\mathbf{v}}$ of muons is also along these surfaces. Yet another way of stating this result is that optimal convergence is achieved if the surfaces of constant gas density are oriented at the Lorentz angle with respect to the electric field. A picture of how this leads to transverse convergence of the muon beam is shown in Fig. 3.

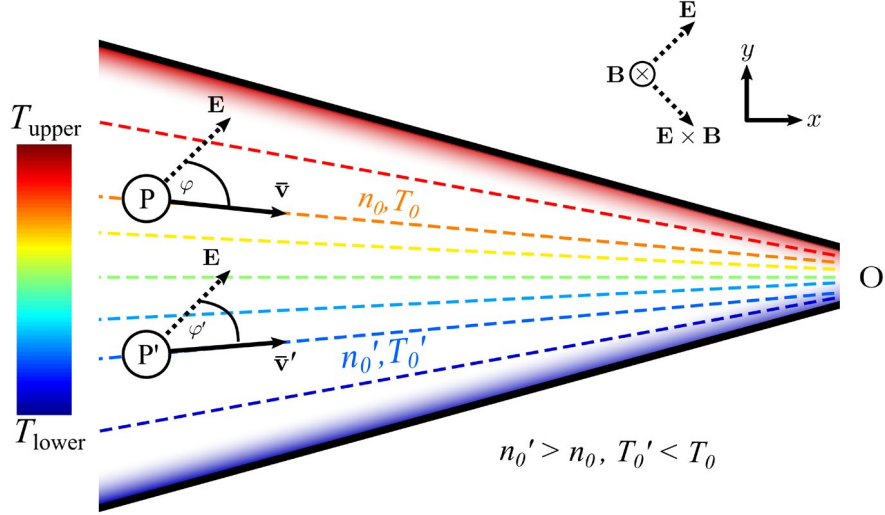

Fig.3 Schematic diagram portraying transverse compression of a muon beam subject to crossed electric and magnetic fields, and in the presence of a gradient in gas density $n_0$, achieved by maintaining upper and lower bounding surfaces at different temperatures. Maximum compression is achieved when the average muon velocity $\bar{v}$ is directed at the local Lorentz angle φ, effectively along a surface of constant gas density. At two different points, P′ and P, where the gas densities and Lorentz angles are $n_0'$, $n_0$ and φ′, φ respectively, with $n_0' > n_0$ and φ′ < φ, the corresponding muon velocities $\bar{v}'$ and $\bar{v}$ converge toward the focus point O.

This concludes an outline of the theoretical framework used for calculation of transport properties needed to analyse the muon beam compression. In what follows we describe how to use this theory in combination with tabulated swarm experimental data in order to furnish experimental parameters for the muon beam experiment.

## 6. Experimental parameters for optimal transverse compression of muon beams

### 6.1 Outline of the procedure

The optimal configuration described above shown in Fig. 3 may be achieved by specifying the experimental parameters through the following procedure:

- Firstly, the direction from the point P in question to the focus point O determines the required Lorentz angle φ at P. It is emphasised that this is a prescribed, independent variable, along with values of the fields E and B and the gas pressure $p_0$.

- Secondly, using Eqns (29) and (30) we obtain following expression for the average speed of muons:

$$\overline{v} = \frac{E}{B} \sin\varphi \qquad (36)$$

This could, if desired, be inserted in Eqn (15) to calculate the corresponding mean energy $\overline{\varepsilon}$ in the C.M. of the muons and the helium gas atoms, but this is not needed here.

- Thirdly, we introduce the *reduced mobility* coefficient,

$$\mathcal{K} = n_0 K / n_L \qquad (37)$$

where $n_L$ = 2.69 x $10^{25}$ m$^{-3}$ is the number density of an ideal gas under standard conditions of pressure and temperature (Loschmidt's number), and rewrite Eqn (30) as

$$\tan\varphi = \frac{n_L \mathcal{K} B}{n_0}$$

Rearrangement then gives gas number density corresponding to the prescribed value of the Lorentz angle:

$$\frac{n_0}{n_L} = \frac{\mathcal{K} B}{\tan\varphi} \qquad (38)$$

The right hand side can be evaluated provided that the reduced mobility of muons in the gas is known. This is equivalent to requiring information about the scattering cross section, since by Eqns (1), (23) and (37), we have

$$\mathcal{K} = \frac{q}{n_L \sqrt{2\mu_r \overline{\varepsilon}} \, \sigma_m(\overline{\varepsilon})} \qquad (39)$$

- Using the value of $n_0$ obtained from (38), and the experimentally specified gas pressure $p_0$, we may then calculate the temperature $T_0$ of the gas along a surface oriented at any angle φ. In particular, we can find the temperatures of the upper and bounding surfaces, which correspond to the minimum and maximum angles respectively.

*6.2 Determination of the muon reduced mobility by the method of aliasing*

The analysis as outlined above is complete except for one vital consideration: data on the mobility of muons in helium gas are generally not available, and the right hand side of the all-important Eqn (38) therefore cannot be evaluated. A similar obstacle arises in simulation techniques, which require information on ($\mu^+$, He) scattering cross sections. Such problems may be obviated by using an approximation called aliasing[10]. Here one adapts the known properties of some other particles in helium, thought to mimic muons, in order to furnish the otherwise unknown data. Following Taqqu[7] and Belosevic[27] we alias ($\mu^+$, $He$) by ($H^+$, $He$) and assume that $\sigma_m(\bar{\varepsilon})$ is the same in both systems for the same value of C.M. mean energy $\bar{\varepsilon}$. With this assumption, it follows from Eqn (39) that the mobilities in the respective systems are simply proportional, i.e.,

$$K(\mu^+, He) = sK(H^+, He) \tag{40}$$

where

$$s = \sqrt{\frac{m_{H^+}(m_{\mu^+} + m_{He})}{m_{\mu^+}(m_{H^+} + m_{He})}} = \sqrt{\frac{36}{5}} \approx 2.7 \tag{41}$$

is a mass scaling factor.

Muon mobilities in helium gas then follow by substituting known swarm experimental data[8,9] for $K(H^+, He)$ in the right hand side of (40). However, as follows from (39), the mobility coefficients on both sides of Eqn (40) must correspond to the same value of $\bar{\varepsilon}$, i.e., we must select those values of $K(H^+, He)$ from tabulated swarm data which correspond to the actual mean C.M. energy $\bar{\varepsilon}$ which occurs in the ($\mu^+$, $He$) beam compression scenario under consideration. This would require solving Eqn (15) for $\bar{\varepsilon}$ with values of $\bar{v}$ specified by (36). If inelastic processes are negligible, then $\Omega \approx 0$ and, assuming the thermal motion of the neutral gas atoms to be also negligible, then Eqn (15) gives immediately $\bar{\varepsilon} \approx \frac{1}{2}m_0\bar{v}^2$. However, things are not so simple if inelastic processes are important, for two reasons:

- $\Omega$ given by Eqn (13) involves inelastic cross sections $\sigma_I(\bar{\varepsilon})$, which would have to be aliased, in addition to $\sigma_m(\bar{\varepsilon})$; and
- Eqn (15) is then rather more difficult to solve for $\bar{\varepsilon}$.

Fortunately these steps can be bypassed using the simple procedure discussed below.

*6.3 Mean energy vs mean velocity as the independent variable*

While the C.M. mean energy $\bar{\varepsilon}$ is common to both sides of Eqn (40), we have, as noted after Eqn (15), that $\bar{v}$ and $\bar{\varepsilon}$ are in one-to-one correspondence. Therefore the mobilities on both sides of (40) can be regarded as functions of $\bar{v}$, instead of $\bar{\varepsilon}$, something which greatly simplifies the calculations. As explained in the Appendix, it is then a straightforward matter to convert tabulated data of $\mathcal{K}$ vs reduced electric field in a swarm experiment to tables of $\mathcal{K}$ vs $\bar{v}$.

*6.4 Summary of results*

Proceeding in this way, we are able to find the required muon mobility data necessary to evaluate the right hand side of Eqn (38), and the results are summarised below in the Table below for three different values of φ. The results are in broad agreement with Taquu's results[7], obtained using an essentially empirical gas density profile in GEANT simulations[4], though without imposing any criterion for optimal convergence.

| φ (deg) | $\bar{v}$ ($10^4$ m/s) | $\frac{1}{2} m_0 \bar{v}^2$ (eV) | $\mathcal{K}$ ($cm^2\ V^{-1} sec^{-1}$) | $n_0/n_L$ ($10^{-2}$) | $T_0$ (K) |
|---|---|---|---|---|---|
| 30 | 1.8 | 6.2 | 157 | 13 | 9 |
| 45 | 2.6 | 14.1 | 189 | 9.5 | 13 |
| 60 | 3.1 | 20.0 | 205 | 5.9 | 21 |

Table 2: Gas temperature and density at various Lorentz angles which optimise transverse compression of muons in helium gas, found from the equations of MTT transport theory, using the ($H^+$, $He$) system to alias required input data, for the same fields, $E = 1.8 \times 10^5$ V/m and $B = 5$ T and gas pressure $p_0 = 4.6\ mbar$ as used by Taqqu[7]. More details can be found in the Appendix.

**7. Discussion and concluding remarks**

This article employs fluid analysis, of a type well known in low energy gaseous ion and electron transport theory, to provide a bridge between analysis of traditional swarm experiments and current experiments involving low energy muons μ$^+$ in gases. Firstly, we show how this approach furnishes an expression for stopping power familiar in beam physics. The main focus is, however, on determining experimental parameters which optimise compression of a muon beam, using a density gradient in the gas. After discussing longitudinal compression in an electric field, we consider transverse compression of a beam in the presence of both electric and magnetic fields, and establish that optimal compression is achieved when the surfaces of constant gas density are oriented parallel to the local Lorentz angle. An analysis was carried out for a μ⁺ beam in helium gas in crossed electric and magnetic fields, and experimental parameters for optimal transverse compression thus obtained. The results are in broad agreement with Taqqu's results[7] obtained from GEANT simulations.

A problem faced in both simulation procedures and transport analysis is that neither scattering cross sections nor swarm experimental data are generally available for muons. Thus, this article follows Taqqu[7], Belosevic[27] and Robson *et al.*[10] in carrying out the calculations through "aliasing" the ($\mu^+$, *He*) system by ($H^+$, *He*), for which scattering cross sections and experimental swarm data are known. The circumstances surrounding the beam compression described here are, however, somewhat specialised, and a more general discussion of the aliasing technique is warranted.

Note that while this article offers a distinct alternative to modelling muon beams relying on standard simulation packages, by offering a self-contained theory involving a minimum of mathematical and computational analysis, we feel that the two different approaches should be regarded as complementary. However, at the same time it is our view that both should be guided by the criterion established here for optimal transverse compression.

Finally, we note application of fluid transport equations for investigation of beams is not by any means confined to low energy (~ 10 eV) muons, as discussed here. Indeed, the approach has been successfully employed to model the highly relativistic, ultra-high energy (~ 1 GeV) electron beams in compact plasma accelerators. There the fluid approach compares favourably with simulations methods like PIC, in both accuracy and efficiency[28,29].

## Appendix

*Estimating experimental parameters for ($\mu^+$, He) from swarm data*

The procedure used to generate the data in Table 2 may be summarised as follows:

1. Firstly, Eqn (36) with Taqqu's values[7] of the fields, $E = 1.8 \times 10^5$ V/m and $B = 5$ T gives

$$\bar{v} = 3.6 \times 10^4 \sin\varphi \quad m/s \qquad (A1)$$

   where the right hand side can be evaluated for the Lorentz angle φ of interest.

2. Next we convert tabulated swarm experiment data[8,9] for $\mathcal{K}(H^+, He)$ vs reduced electric field to $\mathcal{K}(H^+, He)$ vs $\bar{v}$, and select $\mathcal{K}$ corresponding to the value of $\bar{v}$ given by (A1). Eqn (40) then immediately gives $\mathcal{K}(\mu^+, He)$ for the angle φ in question. The details are shown below.

3. The required gas density at this angle then follows from Eqn (38).

4. Finally, the corresponding gas temperature is found from the ideal gas equation of state, for a gas pressure of p = 4.6 mbar[7].

Step 2, establishing the relationship $\mathcal{K} = \mathcal{K}(\bar{v})$ between the reduced mobility $\mathcal{K}$ and drift velocity $\bar{v}$, from swarm experimental data in an electric field only, requires further discussion. Swarm experiments traditionally analyse motion of ions in an electric field $E$ in a neutral gas of number density $n_0$. Measured reduced ion mobilities $\mathcal{K}$ are usually reported in units of cm$^2$ V$^{-1}$ s$^{-1}$, and tabulated as a function of the reduced field $E/n_0$, the latter being expressed in units of Td = 1 townsend = $10^{-21}$ V m$^2$. The usual relationship $\bar{v} = KE$ between drift velocity and electric field can be expressed in terms of the reduced mobility as

$$\bar{v} = 2.69 \, \mathcal{K}_{cm^2 s^{-1} V^{-1}} (E/n_o)_{Td} \quad m\,s^{-1} \qquad (A2)$$

For convenience in what follows, we shall simply write $\mathcal{K}$ for $\mathcal{K}_{cm^2 s^{-1} V^{-1}}$

Note that $E$ and $n_0$ on the right hand side of (A2) represent the electric field and gas density respectively in the swarm experiment from which mobility data are obtained, *not* the quantities in the muon beam experiment under investigation. The common factor linking the two experiments is the drift velocity $\bar{v}$ which, on the one hand for the swarm experiment is determined by (A2) for given mobility data and, on the other hand, for the beam compression experiment is found from (A1) for a given Lorentz angle φ.

The procedure to find $n_0$ which optimises transverse compression of the muon beam is then as follows:

- From given $\mathcal{K}$ vs $E/n_o$ data[8,9] for ($H^+$, $He$) calculate the corresponding swarm drift velocities $\bar{v}$ using (A2). This effectively generates the function $\mathcal{K}(\bar{v})$.

- Reduced mobilities for the beam compression scheme then follow by evaluating $\mathcal{K}(\bar{v})$ with $\bar{v}$ given by (A1), over a range of Lorentz angles, e.g., φ = $30^0$, $45^0$ and $60^0$ respectively.

- This $\mathcal{K}(H^+, He)$ data can readily be converted to values of $\mathcal{K}(\mu^+, He)$ using Eqn (40) of the text.
- Find the gas density $n_0$ corresponding to these angles from Eqn (38), which becomes

$$\frac{n_0}{n_L} = \frac{\mathcal{K}\,B}{\tan\varphi} \times 10^{-4} \qquad (A3)$$

  for $\mathcal{K} = \mathcal{K}(\mu^+, He)$ data expressed in units of cm$^2$ V$^{-1}$ s$^{-1}$.

- Finally, we can calculate temperatures $T_0$ corresponding to these densities from the ideal gas equation of state, $p_0 = n_0\,k\,T_0$, for a given value of gas pressure, e.g., following Taqqu[7], $p_0 = 4.6$ *millibar*.

The results are summarised in Table 2 of the text.